\documentclass[12pt,epsf]{article}
\usepackage{graphicx}
\usepackage{epsfig}
\begin{document}

\def\a{\alpha}
\def\b{\beta}
\def\c{\varepsilon}
\def\d{\delta}
\def\e{\epsilon}
\def\f{\phi}
\def\g{\gamma}
\def\h{\theta}
\def\k{\kappa}
\def\l{\lambda}
\def\m{\mu}
\def\n{\nu}
\def\p{\psi}
\def\q{\partial}
\def\r{\rho}
\def\s{\sigma}
\def\t{\tau}
\def\u{\upsilon}
\def\v{\varphi}
\def\w{\omega}
\def\x{\xi}
\def\y{\eta}
\def\z{\zeta}
\def\D{\Delta}
\def\G{\Gamma}
\def\H{\Theta}
\def\L{\Lambda}
\def\F{\Phi}
\def\P{\Psi}
\def\S{\Sigma}

\def\o{\over}
\def\beq{\begin{eqnarray}}
\def\eeq{\end{eqnarray}}
\newcommand{\gsim}{ \mathop{}_{\textstyle \sim}^{\textstyle >} }
\newcommand{\lsim}{ \mathop{}_{\textstyle \sim}^{\textstyle <} }
\newcommand{\vev}[1]{ \left\langle {#1} \right\rangle }
\newcommand{\bra}[1]{ \langle {#1} | }
\newcommand{\ket}[1]{ | {#1} \rangle }
\newcommand{\EV}{ {\rm eV} }
\newcommand{\KEV}{ {\rm keV} }
\newcommand{\MEV}{ {\rm MeV} }
\newcommand{\GEV}{ {\rm GeV} }
\newcommand{\TEV}{ {\rm TeV} }
\def\diag{\mathop{\rm diag}\nolimits}
\def\Spin{\mathop{\rm Spin}}
\def\SO{\mathop{\rm SO}}
\def\O{\mathop{\rm O}}
\def\SU{\mathop{\rm SU}}
\def\U{\mathop{\rm U}}
\def\Sp{\mathop{\rm Sp}}
\def\SL{\mathop{\rm SL}}
\def\tr{\mathop{\rm tr}}

\def\IJMP{Int.~J.~Mod.~Phys. }
\def\MPL{Mod.~Phys.~Lett. }
\def\NP{Nucl.~Phys. }
\def\PL{Phys.~Lett. }
\def\PR{Phys.~Rev. }
\def\PRL{Phys.~Rev.~Lett. }
\def\PTP{Prog.~Theor.~Phys. }
\def\ZP{Z.~Phys. }


\baselineskip 0.7cm

\begin{titlepage}

\begin{flushright}
UT-07-38 \\
IPMU 07-0015
\end{flushright}

\vskip 1.35cm
\begin{center}
{\large \bf
    Determining the mass for an ultralight gravitino at LHC
}
\vskip 1.2cm
K. Hamaguchi${}^{1}$, S. Shirai${}^{1}$ and T. T. Yanagida${}^{1,2}$
\vskip 0.4cm

${}^1$
{\it  Department of Physics, University of Tokyo,\\
     Tokyo 113-0033, Japan}

\vskip 0.4 cm

${}^2$
{\it Institute for the Physics and Mathematics of the Universe, University of Tokyo, \\
Chiba 277-8568, Japan}

\vskip 3.5cm

\abstract{In supersymmetric (SUSY) models with the gravitino being the lightest SUSY particle (LSP), the SUSY breaking scale (i.e., the gravitino mass) could be determined by measuring the lifetime of the next-to-lightest SUSY particle (NLSP). However, for an ultralight gravitino of mass of ${\cal O}(1)$~eV, which is favored cosmologically, the determination of the SUSY breaking scale, or the gravitino mass, is difficult because the NLSP decay length is too short to be measured directly. Recently we proposed a new determination of the gravitino mass by measuring a branching fraction of two decay modes of sleptons. In this paper, we investigate the prospects for determining the gravitino mass at LHC. For demonstration we take some explicit gauge-mediation models and  
show that the gravitino mass can be determined with an accuracy of  
a few 10\% for an integrated luminosity $10-100$~ fb$^{-1}$.}
\end{center}
\end{titlepage}

\setcounter{page}{2}

\section{Introduction}

The presence of a gravitino is the most fundamental prediction in supergravity~\cite{SUGRA} 
and its mass $m_{3/2}$ is an important parameter to determine 
the supersymmetry (SUSY)-breaking scale. The gravitino mass is predicted in a wide-range region,  
$m_{3/2}=0.1~{\rm eV}~-~100$ TeV, depending on mediation mechanisms of the SUSY breaking to
the SUSY standard-model (SSM) sector. The lowest mass region, $m_{3/2}=0.1~{\rm eV} - 10$ eV, is very
interesting in particular, since there is no astrophysical and cosmological gravitino problem at all 
in this mass region~\cite{Pagels:1981ke+Viel:2005qj}. 
If it is the case, the gravitino is the lightest SUSY particle (LSP), 
and the next-to-lightest SUSY particle (NLSP) decays into the gravitino inside defectors if it is produced at collider experiments.
Therefore, we may have a signal for the NLSP decay into 
the gravitino at future collider experiments such as LHC. Furthermore, we may determine the gravitino mass 
from the mass and lifetime of the NLSP. However, for a light gravitino of mass $m_{3/2}\lsim 10$ eV
the NLSP lifetime is very short; for instance, the decay length is given by
$c\tau_{\rm NLSP} \simeq 0.55~\mu\mathrm{m}
 (m_{3/2}/1~\mathrm{eV})^2 (m_{\rm NLSP}/200~\mathrm{GeV})^{-5}$ for a slepton NLSP, which is
 difficult to measure at collider experiments.

In a recent article~\cite{Hamaguchi:2007ge}, 
we proposed a new method to determine the gravitino mass (i.e., the SUSY-breaking scale)
for a very light gravitino, by
comparing the branching fractions of two decay modes of sleptons, 
${\tilde \ell}_1\rightarrow {\tilde \tau}_1 +\tau +\ell$ and
${\tilde \ell}_1\rightarrow \ell + {\tilde G}_{3/2}$, instead of using
the lifetime of the NLSP. (In this proposal we assume that the NLSP is the lighter stau $\tilde{\tau}_1$. 
$\tilde{\ell}_1$ denotes the lighter smuon or selectron, $\ell$ is $\mu$ or $e$, and
$\tilde{G}_{3/2}$ is the gravitino.) However, we did not
examine if this method works at LHC. The purpose of this paper is to show that the above
method to determine the gravitino mass is indeed effective at LHC 
for a certain parameter
region of the SUSY-particle spectrum.
To demonstrate our point, we adopt simple gauge-mediated SUSY breaking (GMSB) models~\cite{Giudice:1998bp} in this paper. However, 
our mechanism for the determination of the gravitino mass is applicable to any gauge-mediation 
model as long as sleptons are the lightest next to the gravitino LSP and  $m_{3/2}\lsim 10$~eV.

\section{Measurement of the branching fraction for the slepton decays}

In this section we explain the basic idea of our method. 
We assume that the NLSP is the lighter stau $\tilde{\tau}_1$, 
and the lighter smuon and selectron  
(collectively denoted by $\tilde{\ell}_1$) are 
heavier than the stau but lighter than the lightest 
neutralino $\tilde{\chi}^0_1$, $m_{\tilde{\tau}_1} < m_{\tilde{\ell}_1} < m_{\tilde{\chi}^0_1}$.
In this case, $\tilde{\ell}_1$ have two dominant decay modes.
One is the decay into the gravitino, and its decay rate is related to the gravitino mass:
\begin{eqnarray}
\Gamma_{\rm 2-body} \;=\; \frac{m_{\tilde{\ell}_1}^5}{48 \pi M_P^2 m_{3/2}^2}
\;=\; 0.035 {\rm eV} \left( \frac{m_{\tilde{\ell}_1}}{200 ~{\rm GeV}} \right)^5 
 \left( \frac{m_{3/2}}{1 ~{\rm eV}} \right)^{-2}\,,
\label{eq:Gamma2}
\end{eqnarray}
where $M_P =2.44 \times 10^{18}$~GeV is the reduced Planck mass.
The other decay mode is the three-body decay into the lighter stau,
$\tilde{\ell}_1\to \tilde{\tau}_1^{\pm} \tau^{\mp} \ell$. As pointed out in the previous work~\cite{Hamaguchi:2007ge}, if one can observe both of these decay modes, the gravitino mass can be determined.
This is because the gravitino mass is written as
\begin{equation}
m_{3/2}^2 = \frac{m_{\tilde{\ell}_1}^5}{48 \pi M_P^2}
\left(\frac{\Gamma_{\rm 3-body}}{\Gamma_{\rm 2-body}}\right)\frac{1}{\Gamma_{\rm 3-body}}\,, \label{eq:Gmass}
\end{equation}
and $\Gamma_{\rm 3-body}$ is calculable once relevant SUSY particles' masses are known. 
Thus, we may derive the gravitino mass $m_{3/2}$ 
by measuring the branching fraction $\Gamma_{\rm 3-body}/\Gamma_{\rm 2-body}$. 

\begin{figure}[t!]
\hspace{-3mm}
\begin{tabular}{cc}
\begin{minipage}{0.5\hsize}
\begin{center}
{\epsfig{file=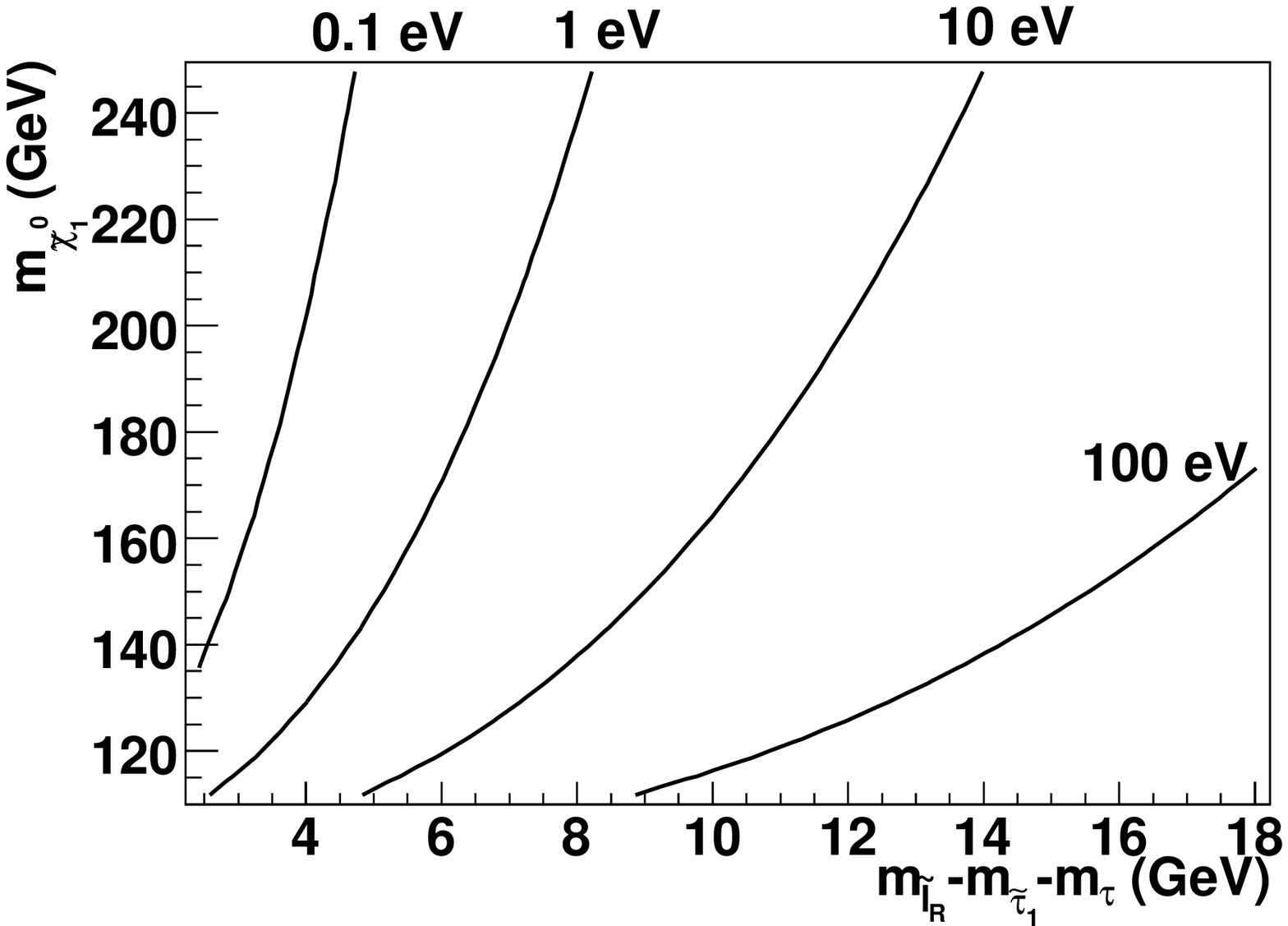, clip,width=9cm}}
\end{center}
\end{minipage}
\begin{minipage}{0.5\hsize}
\begin{center}
{\epsfig{file=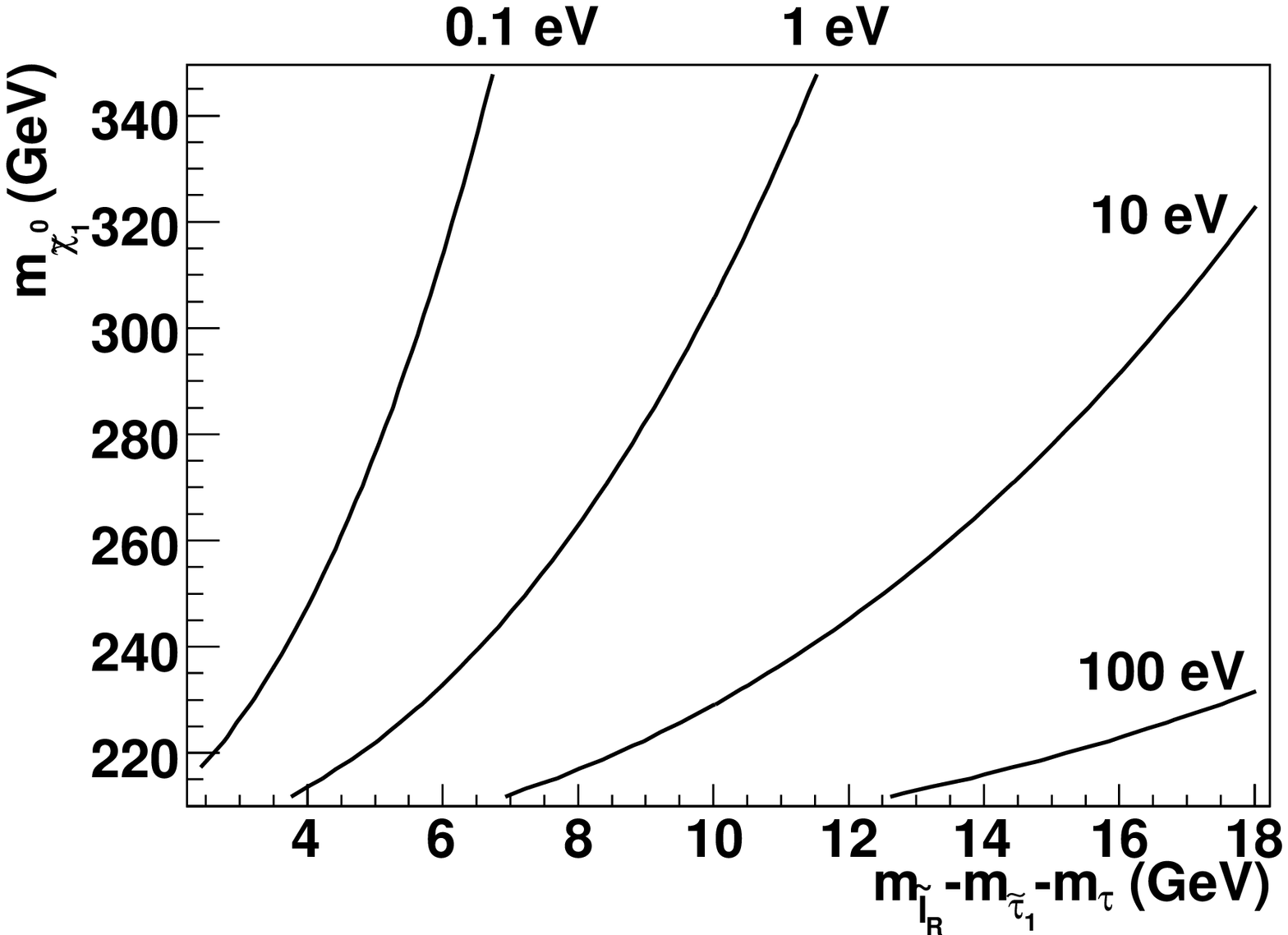, clip,width=9cm}}
\end{center}
\end{minipage}
\end{tabular}
\caption{The contour plot of $\Gamma_{\rm 3-body}$. The left-side figure is for $m_{\tilde{\ell}_R} = 100$~GeV, and the right-side is for $m_{\tilde{\ell}_R} = 200$~GeV. The contour parameter is (0.1, 1, 10, 100) eV from left to right.}
\label{fig:rate}
\end{figure}

Here we should comment on the measurement of the three-body decay rate $\Gamma_{\rm 3-body}$. In principle, $\Gamma_{\rm 3-body}$ depends on various SUSY parameters. However, in most of GMSB models with a light gravitino ($m_{3/2}={\cal O}(1)~{\rm eV}$), the lighter stau, smuon and selectron are approximately right handed sleptons, $\tilde{\tau}_1\sim \tilde{\tau}_R, \tilde{\ell}_1\sim \tilde{\ell}_R$, and the lightest neutralino is almost bino, $\tilde{\chi}^0_1\sim \tilde{B}$. 
Therefore, we assume it is the case in the following discussion.
Under those approximations, the three-body decay rate is given 
by~\cite{Ambrosanio:1997bq}
\begin{eqnarray}
\Gamma_{\rm 3-body} &\simeq& m_{\tilde{\ell}_R}\frac{1}{8\pi}\frac{\alpha_{\rm EM}}{\cos^2\theta_W}\int_{0}^{1-(r_{\tau} + r_{\tilde{\tau}})^2}dx 
x (1-x+r^2_{\tau}-r^2_{\tilde{\tau}}) \left( 1-x  +r^2_{\tilde{B}}  \right) \nonumber \\
&&\times \frac{x\sqrt{(1-x)^2  + r_{\tau}^2 + r_{\tilde{\tau}}^2 -2(1-x)r_{\tau}^2 -2(1-x)r_{\tilde{\tau}}^2-2r_{\tau}^2 r_{\tilde{\tau}}^2 }}{ (1-x )^2(r_{\tilde{B}}^2-1+x)^2},
 \label{eq:3body}
\end{eqnarray}
where
$r_{\tilde{B}} = m_{\tilde{\chi}^0_1}/m_{\tilde{\ell}_R}, r_{\tilde{\tau}} = m_{\tilde{\tau}_1}/m_{\tilde{\ell}_R}, r_{\tau} = m_{\tau}/m_{\tilde{\ell}_R}$, and the lepton masses $m_\mu$ and $m_e$ have been neglected. 
Here, we have defined $\Gamma_{\rm 3-body} = \Gamma(\tilde{\ell}_R\rightarrow \tilde{\tau}_1^+\tau^-\ell) + \Gamma(\tilde{\ell}_R\rightarrow \tilde{\tau}_1^-\tau^+\ell)$.
We have checked that the approximation Eq.(\ref{eq:3body}) is quite good and 
we can reproduce the true $\Gamma_{\rm 3-body}$ with an accuracy of factor 30\% via the approximation Eq.(\ref{eq:3body})
for simple GMSB models explained in the next section.
Therefore, by measuring the three masses $m_{\tilde{\chi}^0_1}$, $m_{\tilde{\ell}_R}$, and $m_{\tilde{\tau}_1}$, one can estimate the 
$\Gamma_{\rm 3-body}$ with a good accuracy.
Contour plots of $\Gamma_{\rm 3-body}$ calculated by Eq.(\ref{eq:3body}) is shown in Fig.{\ref{fig:rate}}.
Comparing them with Eq.(\ref{eq:Gamma2}), one can see that the two-body decay rates $\Gamma_{\rm 2-body}$  are comparable to the three-body decay rates $\Gamma_{\rm 3-body}$ in a certain parameter space.

Let us consider how we can measure the branching fraction 
$\Gamma_{\rm 3-body}/\Gamma_{\rm 2-body}$ at LHC.
Comparing the three-body decay
\begin{eqnarray}
\tilde{\ell}_{R} \;\rightarrow\; \ell+\tau (\rm soft)+\tilde{\tau}_1 \;\rightarrow\;
 \ell+ \tau (\rm soft) + \tau+ \tilde{G}_{3/2},
\end{eqnarray}
with the two-body decay
\begin{eqnarray}
\tilde{\ell}_{R} \;\rightarrow\; \ell+ \tilde{G}_{3/2},
\end{eqnarray}
we can see that the three-body decay accompanies a hard tau.
Thus,
as $\Gamma_{\rm 3-body}/\Gamma_{\rm 2-body}$ is larger, 
the number of taus produced in SUSY-like events becomes larger.
Therefore, one may think that the branching fraction can be estimated by counting the excess of 
the number of hard taus 
in SUSY-like events. 
However, this method is troublesome because of
the difficulty of tau-identification and
the enormous backgrounds.

\begin{figure}[t!]
\begin{center}
\epsfig{file=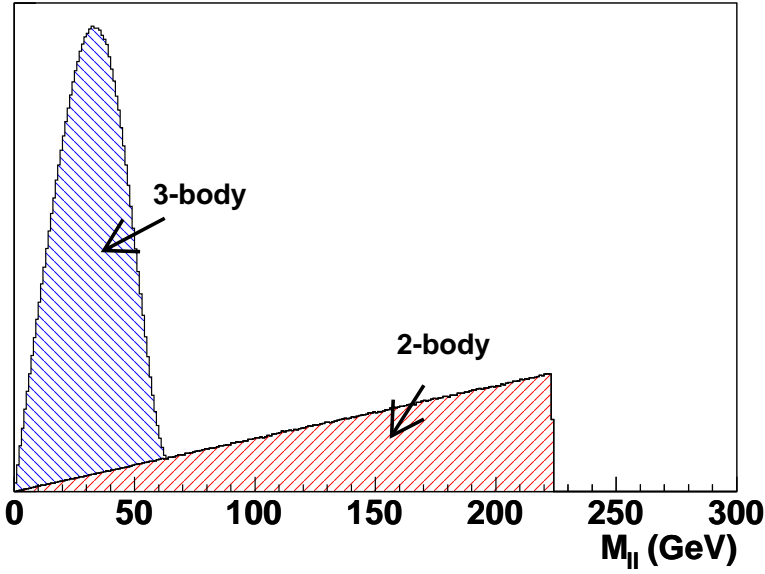, scale =0.6}
\caption{The $M_{\ell^+ \ell^-}$ distribution. Here we set $m_{\tilde{\tau}_1} =190~$GeV, $m_{\tilde{\ell}_R} = 200~$GeV,$m_{\tilde{B}} = 300~$GeV and $m_{3/2}=1$~eV. The $M_{\ell^+ \ell^-}$ distribution from the 2-body decay have an endpoint at $M^{\max}_2 = \sqrt{m_{\tilde{B}}^2 - m_{\tilde{\ell}_R}^2} = 224~$GeV, and 3-body at $M^{\max}_3 =\sqrt{m_{\tilde{B}}^2 - m_{\tilde{\ell}_R}^2}\sqrt{1-(m_{\tau} + m_{\tilde{\tau}_1})^2/m_{\tilde{\ell}_R}^2} = 63~$GeV. In this case, $\Gamma_{\rm 3-body}/\Gamma_{\rm 2-body} = 1.23$. }
\label{fig:idea}
\end{center}
\end{figure}

Here, we propose an alternative experimental method to measure the branching fraction
$\Gamma_{\rm 3-body}/\Gamma_{\rm 2-body}$.
At LHC, $\tilde{\ell}_R$ are likely to be produced through $\tilde{\chi}^0_1$'s decays.
Thus, we consider the two decay chains, $\tilde{\chi}^0_1 \rightarrow \ell^{\pm}\tilde{\ell}_R^{\mp} \rightarrow
\ell^{\pm} {\ell}^{\mp} \tilde{G}_{3/2}$ and
$\tilde{\chi}^0_1 \rightarrow \ell^{\pm}\tilde{\ell}_R^{\mp} \rightarrow
\ell^{\pm} {\ell}^{\mp} \tau\tilde{\tau}_1$.
The dilepton invariant mass from the former chain has distribution with a sharp edge at $M_{\ell^+ \ell^-} =\sqrt{m_{\tilde{\chi}^0_1}^2 - m_{\tilde{\ell}_R}^2}$. On the other hand, for the latter case the dilepton mass distribution has a peak which has an endpoint at $M_{\ell^+ \ell^-} = \sqrt{m_{\tilde{\chi}^0_1}^2 - m_{\tilde{\ell}_R}^2}\sqrt{1-(m_{\tau} + m_{\tilde{\tau}_1})^2/m_{\tilde{\ell}_R}^2}$.
An example of the dilepton mass distribution is shown Fig.\ref{fig:idea}.
From the ratio of these two peaks' areas, we estimate $\Gamma_{\rm 3-body}/\Gamma_{\rm 2-body}$. 

We should note that the latter endpoint is crucial to determine the mass difference $\Delta m = m_{\tilde{\ell}_R} - (m_{\tau} + m_{\tilde{\tau}_1})$, which is one of the most important parameters
 for the calculation of $\Gamma_{\rm 3-body}$
(see Fig.~\ref{fig:rate}).
We also note that there are little background for the signal by virtue of the flavor subtraction technique, 
$e^+e^- + \mu^+\mu^- - e^{\pm}\mu^{\mp}$, where each dilepton represents the final-state one and the sign between each dilepton event defines if the event 
is added or subtracted when booked in the histogram.

\section{Gauge-mediation models}
\label{sec:models}

We consider a simple gauge-mediation model, where
a SUSY breaking field $S$ couples to $N$ pairs of
messenger chiral superfields, $\psi$ and $\bar{\psi}$, 
which transform as
${\bf 5}$ and ${\bf 5}^*$ under the $SU(5)_{\rm GUT}$:
$W = k\psi\bar{\psi}S$. The $S$ field develops a vacuum expectation value $k\langle S \rangle = M + \theta^2 F$, where $M$ is the messenger mass. 
With these conditions the low-energy spectrum of the SUSY particles including
the gravitino mass are determined by 6 parameters, $\Lambda=F/M$, $M$, $N$, 
$\tan\beta$, ${\rm sgn}(\mu) = \pm 1$, and $C_{grav}$~\cite{Giudice:1998bp}. The gaugino masses are generated 
from loop diagrams of the messengers and are given by, at the one-loop level, 
\begin{equation}
m_{a} = \frac{N\alpha_a}{4\pi}\Lambda  ~(a=1,2,3),\label{eq:gaugino_mass}
\end{equation}
where $\Lambda = F/M$ and $\alpha_1=5 \alpha _{\rm EM}/(3 \cos^2\theta_{W})$.
Scalar masses, at the two loop level, are given by
\begin{equation}
m^2_{\phi_i}=2N\Lambda ^2 \sum_a \left(\frac{\alpha_a}{4\pi}\right)^2 C_a (i), \label{eq:scalar_mass}
\end{equation}
where $C_a(i)$ are Casimir invariants for the particle $\phi_i$ 
($C_1(i) = 3Y_i^2/5$).
Here, we have omitted the higher order terms in an expansion in $F/M^2$. The above gaugino and scalar masses are given at the messenger scale, and the physical masses should be obtained by solving the renormalization group equations. Finally, the gravitino mass is given by $m_{3/2} = (C_{grav}/\sqrt{3})(F/M_P)$.

\begin{figure}[t]
\hspace{-3mm}
\begin{tabular}{cc}
\begin{minipage}{1.0\hsize}
\begin{center}
{\epsfig{file=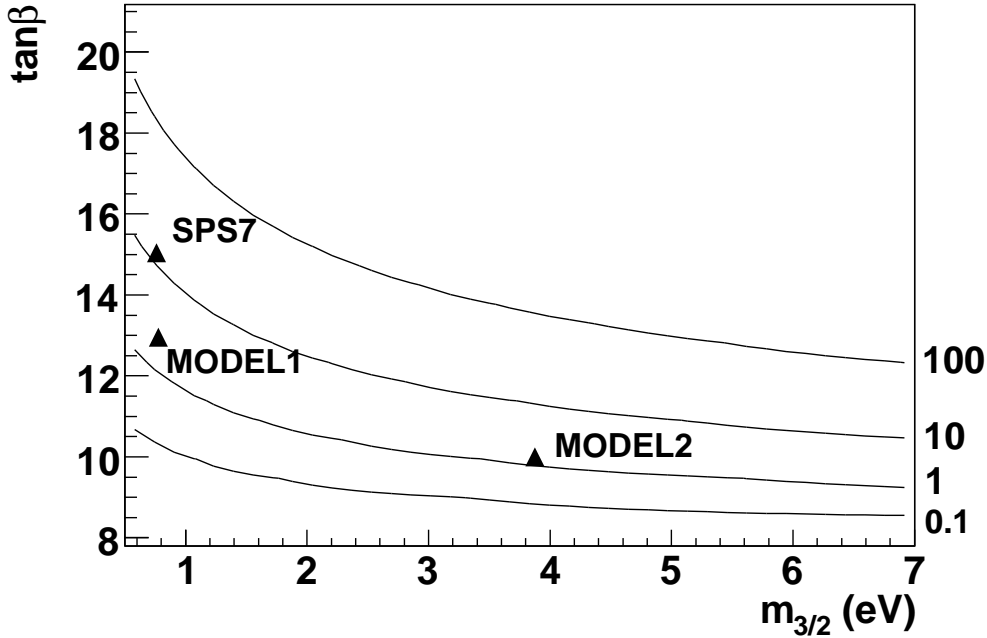, clip,width=14cm}}
\caption{The contour plot of $\Gamma_{\rm 3-body}/\Gamma_{\rm 2-body}$ in the $(m_{3/2},\tan\beta)$ plane.
We set $\Lambda = 40$~TeV, $M = 80$~TeV, $N =3$, ${\rm sgn} (\mu) = +$. In this figure, 
we calculate the mass spectrum and the decay rates by the program 
SOFTSUSY~\cite{Allanach:2001kg} as in Ref.~\cite{Hamaguchi:2007ge}.
The marks represent the model points we discuss in the text. }
\label{fig:N3}
\end{center}
\end{minipage}
\end{tabular}
\end{figure}

For numerical analyses in the next section, we choose a few model points: one is the Snowmass point SPS7~\cite{Allanach:2002nj}, and we also take two other model points to demonstrate the dependence on the model parameters. The GMSB parameters of those models are shown in Table.~\ref{tab:models}.
\begin{table}[t]
\caption{GMSB parameters of the models. }
\begin{center}
\begin{tabular}{lcccccc}
\hline
Point & $\Lambda$~(TeV) & $M$~(TeV) & $N$ & $\tan\beta$ & ${\rm sgn}\mu$ & $C_{grav}$\\ \hline
SPS7   & $40$ & $80$ & $3$ & $15$ & $+$ & $1$\\ 
Model1 & $40$ & $80$ & $3$ & $13$ & $+$ & $1$\\ 
Model2 & $40$ & $80$ & $3$ & $10$ & $+$ & $5$\\ 
\hline
\end{tabular}
\end{center}
\label{tab:models}
\end{table}
In Fig.\ref{fig:N3}, those model points are shown in $(m_{3/2}, \tan\beta)$-plane, together with contour plots of 
$\Gamma_{\rm 3-body}/\Gamma_{\rm 2-body}$.
The full mass spectrum of the models are shown in Table.~\ref{tab:masses}.\footnote{Here, the mass spectrum is calculated by ISAJET 7.67~\cite{ISAJET}, and used in the event generation in the following section.}
\begin{table}[t]
\caption{Mass spectrum }
\begin{center}
\begin{tabular}{lccc|lccc}
\hline
Particle &  SPS7 & Model1 & Model2 & 
Particle & SPS7 & Model1 & Model2 
\\ \hline
$\tilde{g}$ &$952.3$ & $952.1$  &$ 952.4$ &
$\tilde{G}_{3/2}$ & $0.77$~eV& $0.77$~eV  &$3.85$~eV \\ 
$\tilde{u}_L$ &$ 902.1$ & $902.1$ &$ 902.2$ &
$\tilde{u}_R$ & $ 872.8$& $872.7$ &$872.9$ \\
$\tilde{d}_L$ &$ 905.8$ & $905.8$ &$905.8$ & 
$\tilde{d}_R$ &$871.3$ & $871.3$ &$871.4$ \\
$\tilde{b}_2$ &$876.1$& $875.9$ &$875.5$ &
$\tilde{b}_1$ & $ 863.8$& $865.0$ & $866.9$\\
$\tilde{t}_2$ & $ 895.8$& $896.1$ &$896.6$ &
$\tilde{t}_1$ & $ 811.1$& $810.9$ &$810.2$ \\
$\tilde{\nu}$ & $ 251.1$& $251.1$ & $251.1$& 
$\tilde{\nu}_{\tau}$ &$ 250.6$ & $250.7$ &$250.9$ \\
$\tilde{e}_L$ &$267.5$ & $267.5$ &$267.5$ & 
$\tilde{e}_R$ & $129.1$& $129.1$ &$129.0$ \\
$\tilde{\tau}_2$ &$268.8$ & $268.5$ &$268.0$ &
$\tilde{\tau}_1$ &$ 122.3$ & $124.0$ & $126.0$\\
$\tilde{\chi}^0_1$ &$160.1$ & $160.0$ &$159.8$ &
$\tilde{\chi}^0_2$ &$274.9$ & $275.0$ & $275.6$\\
$\tilde{\chi}^0_3$ & $325.0$& $325.6$ &$327.6$ &
$\tilde{\chi}^0_4$ & $388.8$& $389.3$ &$390.9$ \\
$\tilde{\chi}^\pm_1$ &$269.4 $& $269.1$ & $268.9$&
$\tilde{\chi}^\pm_2$ &$391.2$ & $392.0$ &$394.0$ \\
$h^0$ &$113.6$ & $113.4$ & $112.8$&
$H^0$ &$389.7$ & $393.5$ &$400.1$ \\ 
$A$ & $ 389.5$& $393.2$ & $399.5$& 
$H^\pm$ &$ 397.8$ & $401.4$ & $407.5$\\ 
\hline
\end{tabular}
\end{center}
\label{tab:masses}
\end{table}

\section{Determination of the gravitino mass at LHC}

In this section we show that the gravitino mass can indeed be determined at LHC, by taking SPS7, model 1 and 2 as examples. 
In all analyses we use an event generator HERWIG 6.5~\cite{HERWIG6510}.

\begin{figure}[t!]
\begin{center}
{\epsfig{file=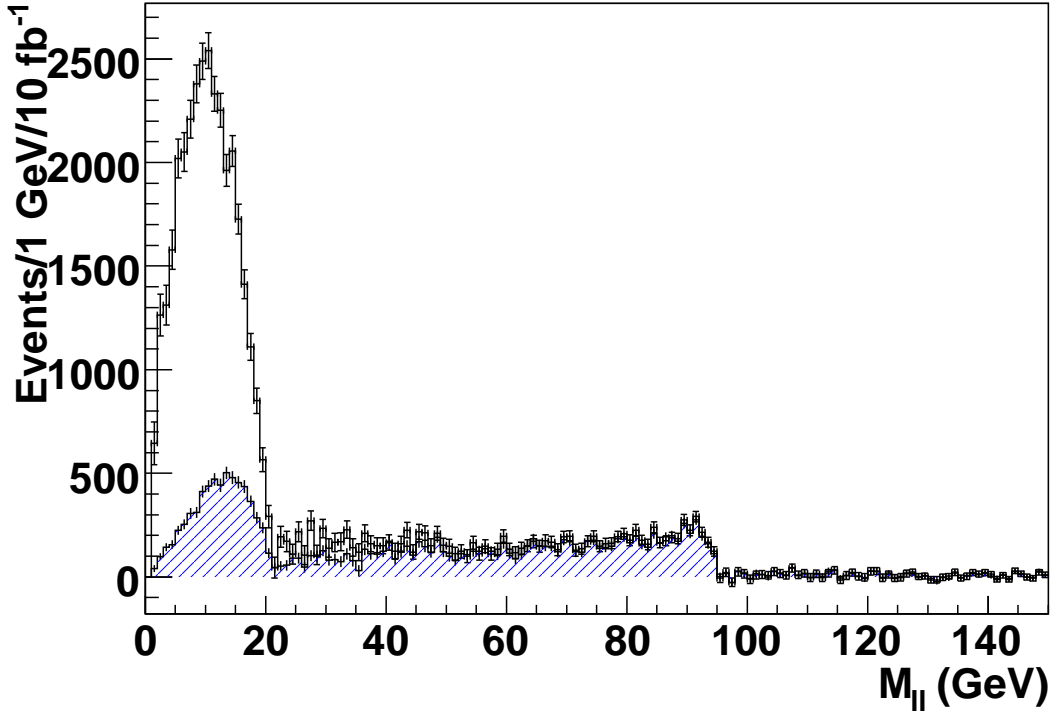, clip,width=15cm}}
\caption{The distribution of $M_{\ell \ell}$.  
As expected, we can see two peaks of endpoints at $M^{\max}_3 = 21.3~{\rm GeV}$ and $M^{\max}_2 = 94.5~{\rm GeV}$.
The hatched histogram shows the distribution after the cuts on the lepton $P_{\mathrm{T}}$.
The peak from the three-body decay is clearly shown in the region of $M_{\ell\ell}\lsim 20~{\rm GeV}$, and 
in $60~{\rm GeV}\lsim M_{\ell \ell}\lsim 90~{\rm GeV}$
one can see the characteristic shape of $M_{\ell \ell}$ distribution from two-body decay.
Note also that in $M_{\ell \ell}\lsim 60~{\rm GeV}$
there is another tiny peak which comes from the decay chain of the next to lightest neutralino,
$\tilde{\chi}^0_2 \rightarrow \ell^{\pm}\tilde{\ell}_R^{\mp} \rightarrow
\ell^{\pm} {\ell}^{\mp} \tilde{\tau}_1\tau$, which has an endpoint at
 $\sqrt{m_{\tilde{\chi}^0_2}^2 - m_{\tilde{\ell}_R}^2}
\sqrt{1-(m_{\tilde{\tau}_1}+m_{\tau})^2/m^2_{\tilde{\ell}_R}} = 54.7~{\rm GeV}$. }
\label{fig:MODEL1_parton}
\end{center}
\end{figure}

Let us first consider the parton-level signatures of signal events.
In Fig.\ref{fig:MODEL1_parton}, the distribution of the 
opposite charge dilepton invariant mass  
$M_{\ell \ell}$ is shown for the model 1.\footnote{It seems that the matrix element of the three-body decay $\tilde{\ell}_R\to \tilde{\tau}_1^{\pm} \tau^{\mp} \ell$ is not implemented in Herwig (i.e., the matrix element is taken to be constant).  Note that the end point of $M_{\ell\ell}$ distribution from the three-body decay and the branching fraction is unchanged even if one takes it into account, and therefore only the shape of the first peak is affected. One may wonder that the reduction factor Eq.~(\ref{eq:R}) due to the soft-lepton cut $P_{\mathrm{T}}>6$~GeV is affected by this approximation, but we have checked that the inclusion of the correct matrix element does not change the result much.}
Here the flavor-subtraction in the final state, $e^+e^-+\mu^+\mu^- - e^{\pm}\mu^{\mp}$, 
is adopted. As expected, one can clearly see the two peaks from two- and three-body decays of the slepton.

However, events are subject, at LHC, to various cuts from experimental constraints and from reducing the backgrounds, and hence the real event distribution may not necessarily follow the ideal one.
In fact, in many cases, the reduction factor
\begin{equation}
R =  \frac{ {\rm \# ~of~dileptons ~with~ the~ cuts} }{  {\rm \# ~of~dileptons ~without~ any~ cuts} }
\label{eq:R}
\end{equation}
tends to be non-negligible. If the reduction factor differs between for two-body and for three-body decay events, the $\Gamma_{\rm 3-body}/\Gamma_{\rm 2-body}$ is affected by the cuts. 

In fact, we require that
\begin{itemize}
\item
the dilepton mass is formed only if one of the two leptons has $P_{\mathrm{T}} \ge 20$ GeV,  $|\eta|<2.5$ and the other has $P_{\mathrm{T}} \ge 6$ GeV,\footnote{We have checked that our method works even if a harder cut $P_{\mathrm T}  > 10$~GeV is taken.} $|\eta|<2.5$,
\end{itemize}
where $P_{\mathrm{T}}$ and $\eta$ denotes the transverse momentum and the pseudorapidity, respectively.
This requirement affects the ratio $\Gamma_{\rm 3-body}/\Gamma_{\rm 2-body}$.
Because of the small difference between $m_{\tilde{\tau}_1}$ and $m_{\tilde{\ell}_R}$, the lepton from the three-body decay is typically soft.
 Therefore, the first peak from the three-body decay shrinks by the $P_{\mathrm T}$ cuts.
The distribution of $M_{\ell \ell}$ after the cuts is also shown in Fig.\ref{fig:MODEL1_parton}, by the hatched histogram. As can be seen, only the events of three-body decay is substantially reduced.

\begin{figure}[t!]
\hspace{-3mm}
\begin{tabular}{cc}
\begin{minipage}{0.45\hsize}
\begin{center}
{\epsfig{file=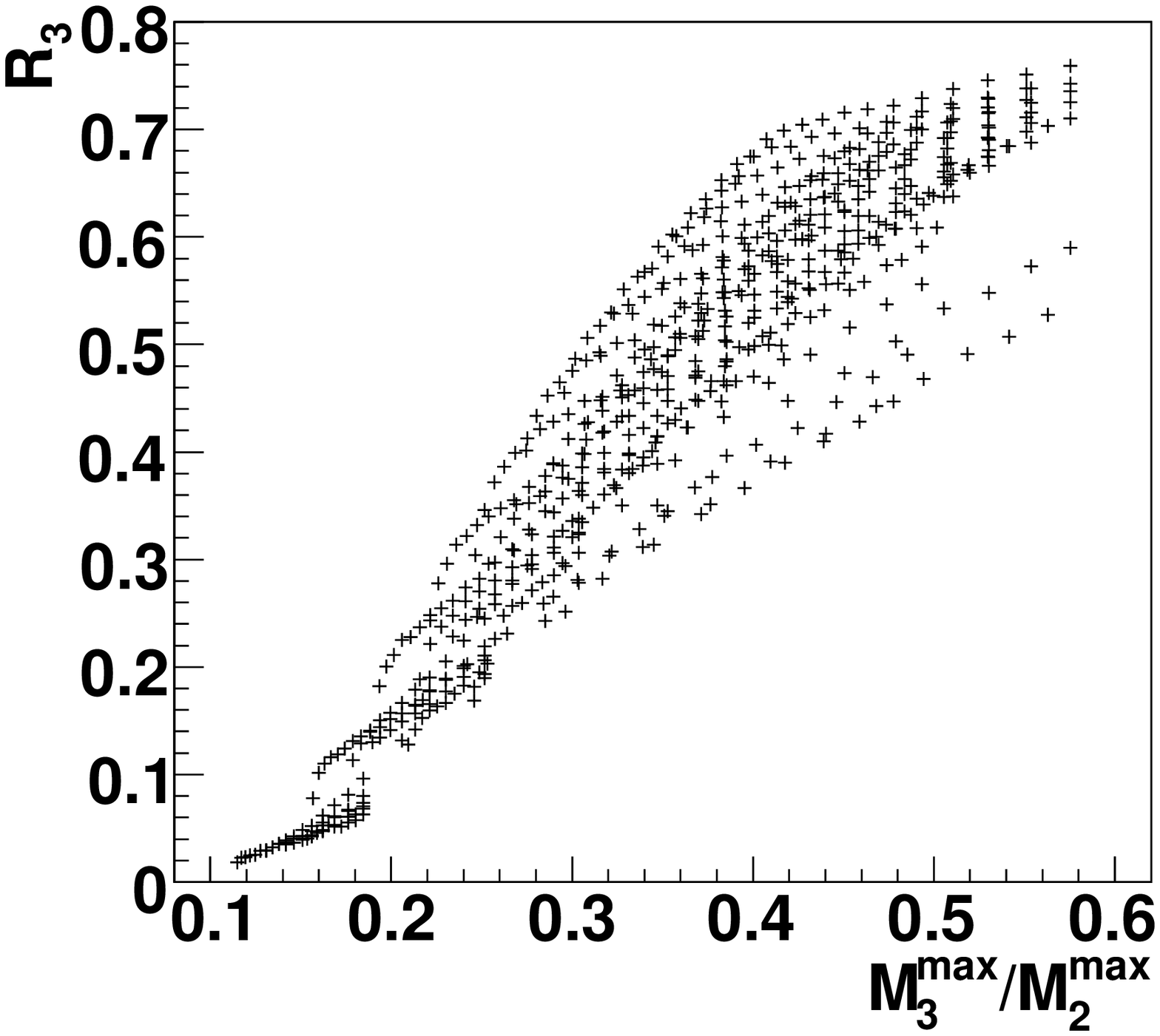, clip,width=7cm}}
\end{center}
\end{minipage}
\begin{minipage}{0.45\hsize}
\begin{center}
{\epsfig{file=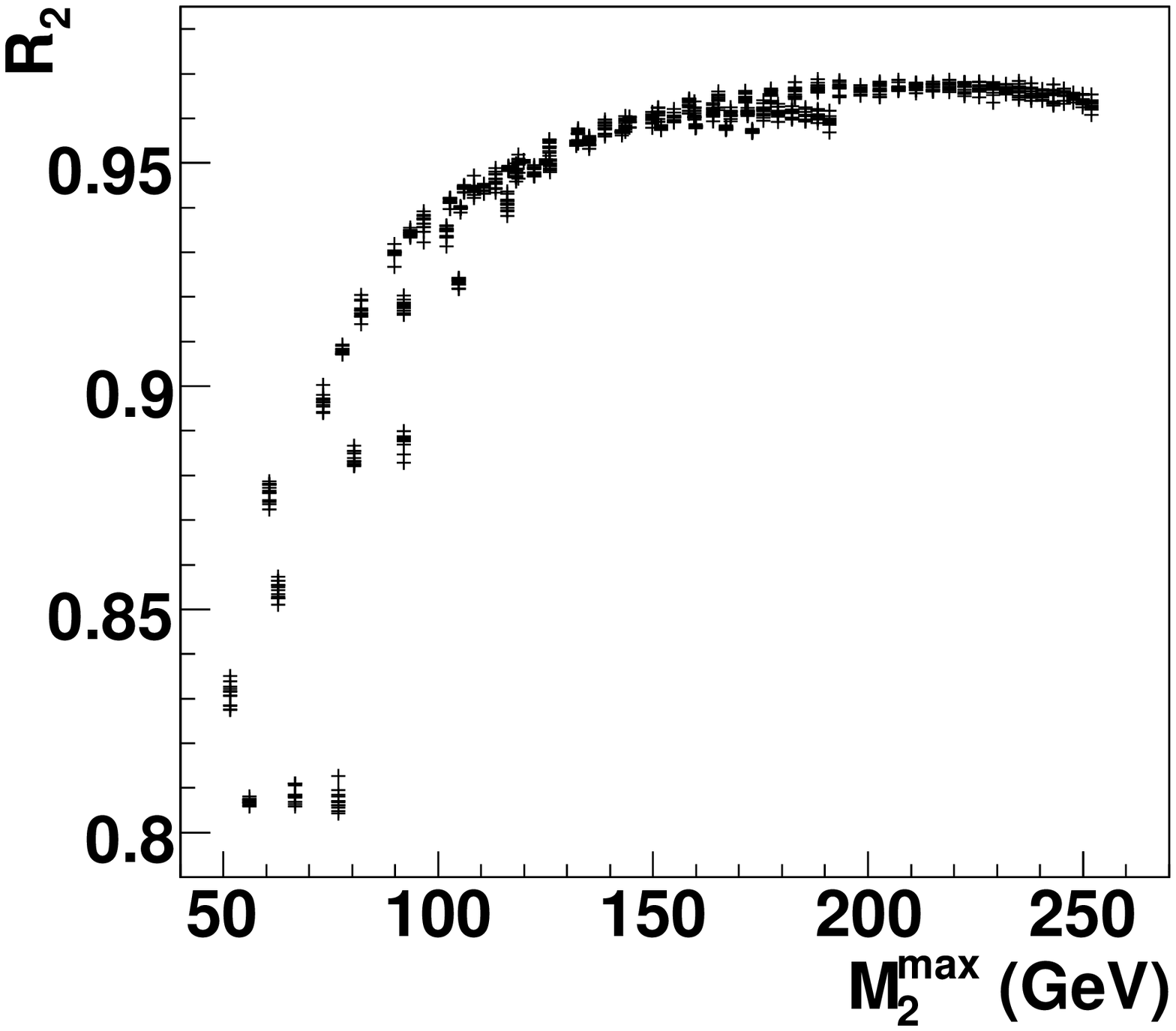, clip,width=7cm}}
\end{center}
\end{minipage}
\end{tabular}
\caption{The correlation between $M^{\max}_3 / M^{\max}_2$ and $R_3$ for three-body decay (left) and
$M^{\max}_2$ and $R_2$ for two-body decay (right).
We have adopted simple GMSB models with parameters ($N$, $\Lambda$, $M$, $\tan\beta$) $=$ 
(\{3, 4, 5\}, $30$ TeV, $60$ TeV, 15) and $=$(\{3, 4, 5\}, $40$ TeV, $80$ TeV, 18).
In each model, we further vary the value of $m_{\tilde{\tau}_1}$ and $m_{\tilde{\ell}_R}$ as free parameters, imposing $m_{\tilde{\chi}^0_1}>m_{\tilde{\ell}_R} + 10~{\rm GeV}$ and $m_{\tilde{\ell}_R} >100~{\rm GeV}$. 
}
\label{fig:reduc}
\end{figure}

We have found that 
there is a correlation between the ratio of two peaks' endpoints $M^{\max}_3/M^{\max}_2 = \sqrt{1-(m_{\tilde{\tau}_1}+m_\tau)^2/m_{\tilde{\ell}_R}^2}$ and the reduction factor $R_3$ of the three-body decay as shown in Fig.\ref{fig:reduc}. Here we have adopted various parameter regions in the simple GMSB model described in Sec.~\ref{sec:models}. We will use this correlation to reproduce the true branching fraction.

Now let us discuss experimental signatures for each models in turn. 
For a detector simulation, we use a package AcerDET-1.0~\cite{RichterWas:2002ch}.
We select events by the following requirements;
\begin{itemize}
\item
at least four jets with $P_{\mathrm{T}} \ge 25$ GeV, where $\tau$-jets are excluded.
\item missing transverse momentum $P_{\mathrm{T,miss}} \ge 100~{\rm GeV}$.
\item $M_\mathrm{eff} \ge 500$ GeV, where
\begin{eqnarray}
M_\mathrm{eff} = \sum_{\mathrm{jets}(\ne \tau)}^{4}
P_\mathrm{T_j} + P_\mathrm{T,miss}\,.
\end{eqnarray}
\item two leptons with $P_{\mathrm{T}} \ge 20$ GeV and $|\eta|<2.5$.
\end{itemize}
We have checked that the first three cuts reduce the two decay chains 
$\tilde{\chi}^0_1 \rightarrow \ell^{\pm}\tilde{\ell}_R^{\mp} \rightarrow
\ell^{\pm} {\ell}^{\mp} \tilde{G}_{3/2}$ and
$\tilde{\chi}^0_1 \rightarrow \ell^{\pm}\tilde{\ell}_R^{\mp} \rightarrow
\ell^{\pm} {\ell}^{\mp} \tau\tilde{\tau}_1$
almost equally and hence 
the ratio of the number of dileptons from the two- and three-body decay chains remains almost unchanged.
The cut of two high $P_{\mathrm T}$ leptons slightly change the ratio, but not very much.\footnote{
In fact, when a pair of sparticle cascades in a SUSY event end up with two decay chains where one is $\cdots\to \tilde{\chi}_1^0\to \tilde{\tau}_1\tau$ and the other is $\cdots\to \tilde{\chi}_1^0\to \tilde{\ell}_R\ell\to \tilde{\tau}_1\tau\ell({\rm soft})\ell$, this event is likely to be cut by this requirement of two high $P_{\mathrm T}$ leptons. This reduces the number of three-body decay events, and decreases the resultant gravitino mass by about 20\%.}
We then form the dilepton invariant mass for opposite charge leptons which satisfy $P_{\mathrm{T}} \ge 20$ GeV,  $|\eta|<2.5$ and $P_{\mathrm{T}} \ge 6$ GeV, $|\eta|<2.5$.\footnote{
In our analysis, we do not take into account of miss-identification of soft leptons.
} As discussed above, this last requirement affect the ratio of number of events, and hence the reduction factor $R$ should be taken into account.

\begin{figure}[t!]
\begin{center}
{\epsfig{file=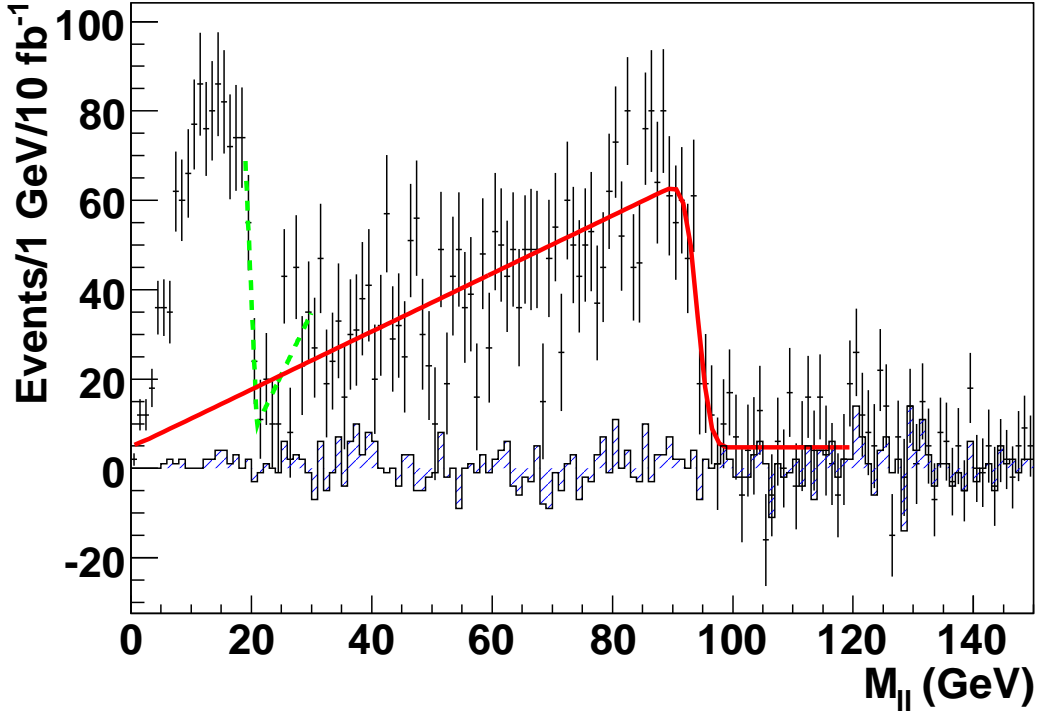, clip,width=15cm}}
\caption{The distribution of $M_{\ell \ell}$ for the model 1.  The hatched histogram is standard model background mainly from $t$-$\bar{t}$ production}
\label{fig:MODEL1_det}
\end{center}
\end{figure}
\subsection{The model 1}
In Fig.\ref{fig:MODEL1_det}, the flavor subtracted ($e^+e^-+\mu^+\mu^- - e^{\pm}\mu^{\mp}$) distribution of the dilepton mass after the cuts is shown for the model 1. 
Here, we have assumed an integrated luminosity ${\cal L} = 10$~fb$^{-1}$. We also show the main background from the $t$-$\bar{t}$ events. We fit the data over $70~{\rm GeV}< M_{\ell \ell}<120~{\rm GeV}$ via a line smeared with a Gaussian, and obtain
\begin{equation}
M^{\max}_2 = 94.1 \pm 0.5~{\rm GeV}.
\end{equation}
To find $M^{\max}_3$, we fit the data with a function
\begin{equation}
h(x) = a(x-M)\theta(-x+M)+bx+c\,, \label{eq:fit}
\end{equation}
over $19~{\rm GeV}<M_{\ell\ell}<30~{\rm GeV}$, where $x=M_{\ell\ell}$.
Then we get
\begin{equation}
M^{\max}_3 = 21.1 \pm 1~{\rm GeV}.
\end{equation}
The estimation of the error is done by 'eye'.
Then, we find that the number of dileptons for the two-body decay is $N_2 = 2876\pm 201$, 
and $N_3 = 958\pm 51$ for the three-body decay.\footnote{
Here, those numbers are obtained by extrapolating the fitted line. It may be modified at lower $M_{\ell\ell}$ region depending on the detector performance (cf. \cite{Gjelsten:2004ki}), but such an effect is small.}$^{,}$\footnote{
The estimation of $N_3$ could be affected by the contamination of the events from the decay of the second lightest neutralino $\tilde{\chi}_0^2$, but its effect is negligible (cf. Fig.~\ref{fig:MODEL1_parton}).}
From Fig.\ref{fig:reduc}, we estimate the reduction factor of the three-body decay is $R_3 = 0.20 \pm 0.05$, and those for the two-body decay $R_2 = 0.90\pm0.02$. 
Here, the errors are systematic.
Then, we find\footnote{Here, the correlation between $N_2$ and $N_3$ is neglected for simplicity, taking those parameters to be independent.}
\begin{equation}
\frac{\Gamma_{\rm 3-body}}{\Gamma_{\rm 2-body}}=\frac{N_3 R_2}{N_2 R_3} 
=(1.50 \pm 0.15) \left( \frac{R_2}{0.90} \right)\left( \frac{R_3}{0.20} \right)^{-1}
\label{eq:model1-ratio}
\end{equation}

Suppose that we know $m_{\tilde{\ell}_R} = 129.1\pm 0.5$~GeV which will be determined from additional observations like $m_{j\ell}$ and $m_{j\ell\ell}$ distributions~\cite{GMSBmasses}.
We can then calculate the approximate $\Gamma_{\rm 3-body}$ from $m_{\tilde{\ell}_R}$, $M^{\max}_2$ and $M^{\max}_3$, by using Eq.(\ref{eq:3body}). In the present case, we obtain
\begin{equation}
\Gamma_{\rm 3-body} = 0.21 ^{+0.09}_{-0.07} ~{\rm eV}  ~~({\rm the~true~value~is }~\Gamma_{\rm 3-body}=0.22~{\rm eV}).
\label{eq:model1-3body}
\end{equation}
Combining Eq.(\ref{eq:model1-ratio}) and (\ref{eq:model1-3body}) we derive the gravitino mass from Eq.(\ref{eq:Gmass}) as
\begin{eqnarray}
m_{3/2} = (0.53 ^{+0.11}_{-0.10})\left( \frac{R_2}{0.90}
\right)^{\frac{1}{2}} \left( \frac{R_3}{0.20} \right)^{-\frac{1}{2}}~{\rm eV}.
\end{eqnarray}
This value is in a good agreement with the expected one $m_{3/2}=0.77$~eV.

\begin{figure}[t!]
\begin{center}
{\epsfig{file=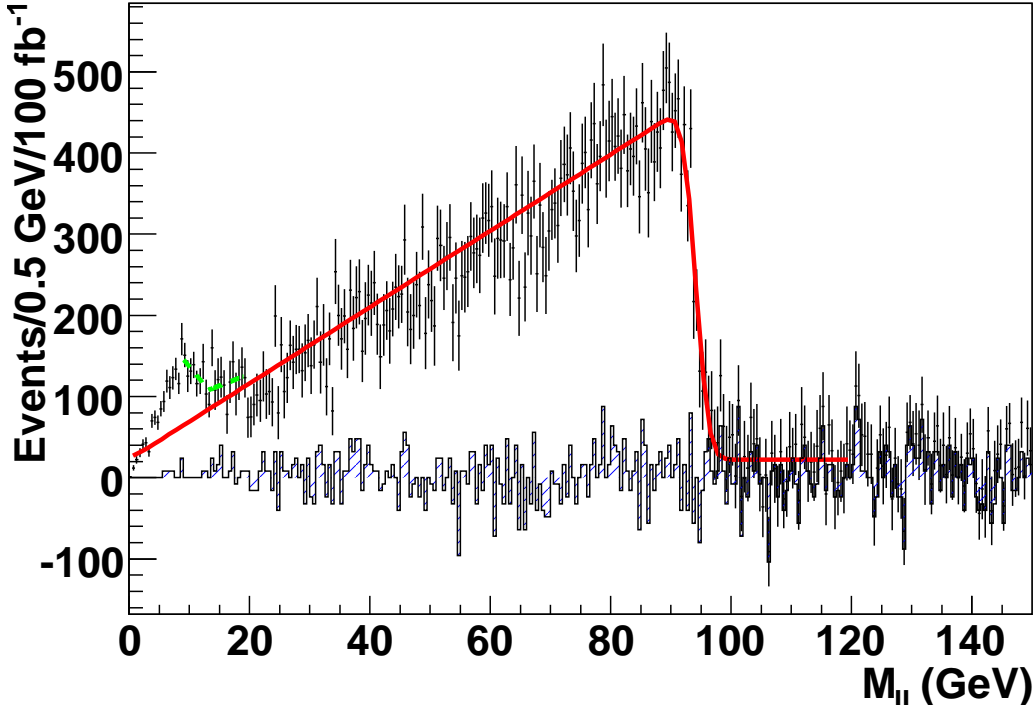, clip,width=15cm}}
\caption{The distribution of $M_{\ell \ell}$ for the model 2.  The hatched histogram is standard model background.
}
\label{fig:MODEL2_det}
\end{center}
\end{figure}
\subsection{The model 2}
In Fig.\ref{fig:MODEL2_det}, the flavor subtracted ($e^+e^-+\mu^+\mu^- - e^{\pm}\mu^{\mp}$) distribution of the dilepton mass after the cuts is shown for the model 2, for an integrated luminosity ${\cal L} = 100$~fb$^{-1}$. We also show the main background from $t$-$\bar{t}$ events.
In this case, an integrated luminosity ${\cal L} = 100$~fb$^{-1}$ would be necessary to determine the gravitino mass. 

We fit the data over $70~{\rm GeV}< M_{\ell \ell}<120~{\rm GeV}$ via a line smeared with a Gaussian and find
$M^{\max}_2 = 94.1 \pm 0.2~{\rm GeV}$.
Also, we fit over $9~{\rm GeV}< M_{\ell \ell}<19~{\rm GeV}$ via $h(x)$ in Eq.(\ref{eq:fit}) and find
$M^{\max}_3 = 13.2 \pm 3~{\rm GeV}$.
Then, we estimate $N_2 = 41627\pm 1066$ and $N_3 = 1561\pm 784$.
We can see $R_3 = 0.04 \pm 0.02 , R_2 = 0.90 \pm 0.02$ from Fig.\ref{fig:reduc}.
Then, we obtain
\begin{equation}
\frac{\Gamma_{\rm 3-body}}{\Gamma_{\rm 2-body}}=\frac{N_3 R_2}{N_2 R_3} 
=(0.85 \pm 0.42) \left( \frac{R_2}{0.90} \right)\left( \frac{R_3}{0.04} \right)^{-1}
\end{equation}
Supposing $m_{\tilde{\ell}_R} = 129.0 \pm 0.5~{\rm GeV}$, we estimate
\beq
\Gamma_{\rm 3-body} =  0.006 ^{+0.021}_{-0.005} {\rm eV}~({\rm the~true~value~is }~\Gamma_{\rm 3-body} = 0.006~{\rm eV}),\,
\eeq
which leads to
\begin{eqnarray}
m_{3/2} = (2.3 ^{+4.5}_{-1.4})\left( \frac{R_2}{0.90}
\right)^{\frac{1}{2}} \left( \frac{R_3}{0.04} \right)^{-\frac{1}{2}}~{\rm eV}.
\end{eqnarray}
The expected value is $m_{3/2} = 3.85$~ eV.

\begin{figure}[t!]
\begin{center}
{\epsfig{file=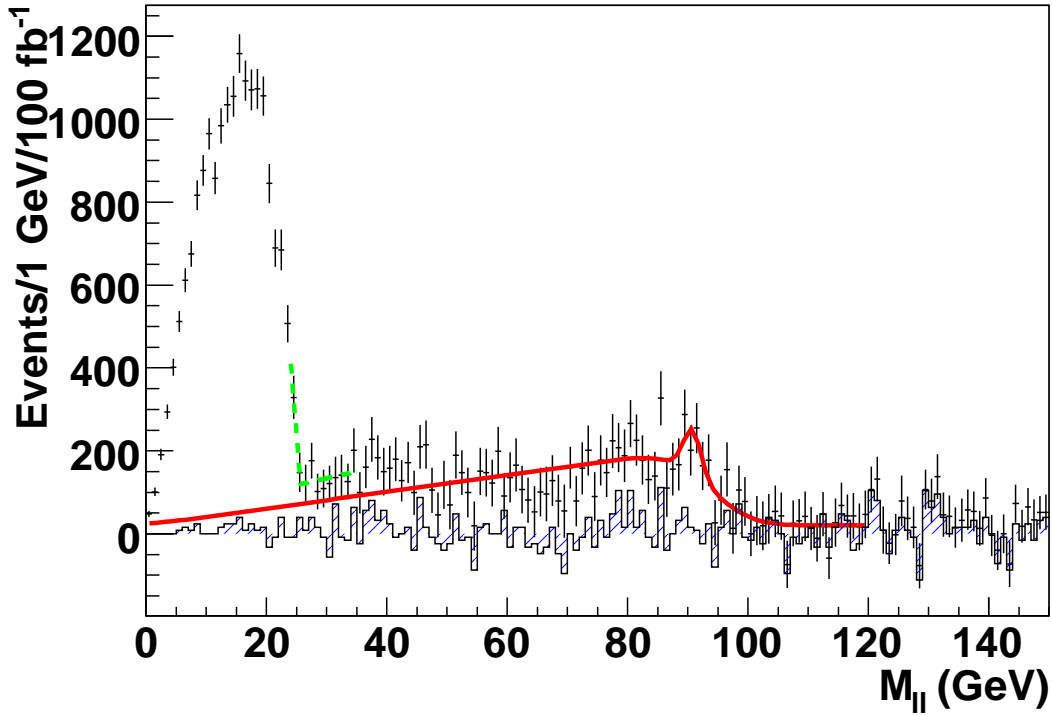, clip,width=15cm}}
\caption{The distribution of $M_{\ell \ell}$ at SPS7.  The hatched histogram is standard model background. The peak around 90 GeV is due to the $Z^0$ decays.
}
\label{fig:SPS7_det}
\end{center}
\end{figure}
\subsection{The SPS7}
In Fig.\ref{fig:SPS7_det}, the flavor subtracted ($e^+e^-+\mu^+\mu^- - e^{\pm}\mu^{\mp}$) distribution of the dilepton mass after the cuts is shown for the SPS7 model point, for an integrated luminosity ${\cal L} = 100$~fb$^{-1}$. We also show the main background from $t$-$\bar{t}$ events.

We fit the data over $70~{\rm GeV}< M_{\ell \ell}<120~{\rm GeV}$ via a line smeared with a Gaussian plus a peak from $Z^0$-boson decays and find
$M^{\max}_2 = 93.5 \pm 1.4~{\rm GeV}$.
Also we fit the data over $24~{\rm GeV}<M_{\ell \ell}<35~{\rm GeV}$ via $h(x)$ in Eq.(\ref{eq:fit}) and find
$M^{\max}_3 = 25.7 \pm 2~{\rm GeV}$.
Then we estimate $N_2 = 8860\pm 1281$ and $N_3 = 17265\pm 498$.
We can see $R_3 = 0.35\pm 0.10, R_2 = 0.90 \pm 0.02$ from Fig.\ref{fig:reduc}.
Thus, we obtain 
\begin{equation}
\frac{\Gamma_{\rm 3-body}}{\Gamma_{\rm 2-body}}=\frac{N_3 R_2}{N_2 R_3} 
=(5.01 \pm 0.84) \left( \frac{R_2}{0.90} \right)\left( \frac{R_3}{0.35} \right)^{-1}
\end{equation}
Assuming that we know $m_{\tilde{\ell}_R} = 129.1\pm0.5~{\rm GeV}$, we estimate
\begin{equation}
\Gamma_{\rm 3-body} = 1.16 ^{+ 1.28}_{-0.59} {\rm eV}~ ({\rm the~true~value~is }~\Gamma_{\rm 3-body} = ~1.11~{\rm eV}),
\end{equation}
which leads to
\begin{eqnarray}
m_{3/2} = (0.41 ^{+0.17}_{-0.12})\left( \frac{R_2}{0.90}
\right)^{\frac{1}{2}} \left( \frac{R_3}{0.35} \right)^{-\frac{1}{2}}~{\rm eV}.
\end{eqnarray}
The expected value is $m_{3/2}=0.77$ ~eV.

\section{Conclusion and Discussion}
We have investigated the prospects for determining the mass of an ultralight gravitino at LHC, by measuring the branching fraction of two decay modes of sleptons. We have performed detailed analyses by taking some specific GMSB models, and demonstrated that the proposed method can indeed work at LHC. Although we have taken simple GMSB models, our method works independently of details of GMSB models, as far as the two decay modes of sleptons are seen and the relevant mass parameters are all known.

So far in this paper, we have assumed that the missing particle is the gravitino LSP and discussed how to determine its mass by measuring the two decay modes of sleptons. Conversely, one may argue that, if the two characteristic peaks of two- and three-body decays of sleptons are simultaneously seen in the dilepton invariant mass distribution, they themselves suggest that the missing particle is not a neutralino but the gravitino LSP, and therefore the underlying model is a GMSB model. One can then perform analyses as presented in this paper and determine the gravitino mass, or equivalently the SUSY breaking scale, which will be one of the most important physics target after the discovery of SUSY.

\section*{Acknowledgement}
We would like to thank the organizers of "LHC visiting program" at KEK, June 2007, especially M.~M.~Nojiri for the tutorials on the tools for high energy physics. We also thank S.~Asai for discussions.
The work by KH was supported by JSPS (18840012).

\end{document}